 \documentclass[final,1p,times]{elsarticle}

\usepackage{graphics}
\usepackage{graphicx}
\usepackage{epsfig}
\usepackage{epstopdf}

\usepackage{amssymb}

\journal{Nuclear Physics B}

\begin{document}

\begin{frontmatter}



\title{Parametrization of light nuclei quasiparticle energy shifts and composition of warm and dense nuclear matter}


\author{ G. R\"opke}

\address{Universit\"at Rostock, Institut f\"ur Physik, 18051 Rostock, Germany}

\begin{abstract}
Correlations and the formation of bound states (nuclei) are essential for the properties of nuclear matter in equilibrium 
as well as in nonequilibrium. In a quantum statistical approach, quasiparticle energies are obtained for the light elements that reflect the influence 
of the medium. We present analytical fits for the quasiparticle energy shifts of light nuclei that can be used in various applications.
This is a prerequisite for the investigation of warm and dense matter that reproduces the nuclear statistical equilibrium and 
virial expansions in the low-density limit as well as relativistic mean field and Brueckner Hartree-Fock approaches near 
saturation density.

\end{abstract}


\end{frontmatter}



 \section{\label{sec:introduction}
 Introduction}
 
 We consider nuclear matter at moderate temperatures, $T \le 20$ MeV, and subsaturation densities, i.\ e.\ baryon densities $n\le n_s \approx 0.16$ fm$^{-3}$. 
The asymmetry given by the proton fraction $Y_p=n_p/n$ is arbitrary, where $n_p$ denotes the total density of protons, and $n_n=n-n_p$ is the total density of neutrons. 
In this regime, correlations between nucleons are important, in particular the formation of bound states, i.e.\ nuclei. 

A general treatment of correlations and cluster formation of the interacting many-nucleon system can be given within a quantum statistical framework \cite{FW}. 
Equilibrium properties such as the equation of state (EOS) as well as transport properties are fundamentally influenced by the composition of nuclear matter. 
The investigation of the properties of nuclear matter in this region of parameter values for $T, n$, and $ Y_p$ (warm nuclear matter) 
is of relevance for astrophysical applications such as supernova explosions, but also for the description of heavy ion collisions (HIC).

Near saturation density, a quasiparticle approach can be used to describe nuclear matter. 
The medium effects result in a self-energy shift of the nucleons that can be incorporated 
in the chemical potentials $\mu_n,\mu_p$ of neutrons and protons, respectively, and in the effective masses $m^*_n, m^*_p$. 
Different semi-empirical approaches such as the Skyrme parametrization \cite{LS} 
or more sophisticated relativistic mean-field (RMF) approaches \cite{Shen} are used to describe the quasiparticle energy shift of nucleons in dense matter, 
in particular in astrophysical applications. 
A more fundamental treatment is possible in quantum statistical approaches such as the Dirac Brueckner Hartree-Fock (DBHF) approximation, see \cite{Klahn}. 
Medium effects become relevant at nucleon density above $\sim 10^{-4}$ fm$^{-3}$. 
However, such single-nucleon quasiparticle approaches fail to describe warm nuclear matter at low densities because correlations, 
in particular bound state formation, are neglected in mean-field theories.

 In the low-density limit, warm nuclear matter can be described as a mixture of different components, the clusters. 
 In equilibrium, a mass action law can be derived, and the composition is determined by nuclear statistical equilibrium (NSE), see \cite{BM}. 
The full account of two-particle correlations ($A=2$) in the equation of state results in the Beth-Uhlenbeck expression for the second virial coefficient \cite{RMS}. 
However,  NSE in which the interaction between the different components is neglected, is possible only below nucleon densities of $\sim 10^{-4}$ fm$^{-3}$. 
 To include interactions on the mean-field level, a generalized Beth-Uhlenbeck formula has been derived in a quantum statistical approach \cite{RMS,SRS} 
that reproduces the exact low-density limit as well as the quasiparticle approach at saturation density. 
In addition to the single-nucleon self-energy shifts, also Pauli blocking has to be considered to describe the mean-field effects for bound states consistently.
 
 The formation of higher clusters ($A>2$) can also be included in the quantum statistical approach \cite{roe84}. 
A cluster-virial expansion \cite{HS} leads to the generalization of the Beth-Uhlenbeck formula for interacting clusters 
that are considered as new components (chemical picture). In particular, the formation of $\alpha$ particles is of importance. 
This has been accounted for  in astrophysical equations of state \cite{LS,Shen}. The contribution of clusters near saturation density,
 however, has to be suppressed. In a simple phenomenological approach, this suppression is modeled in an {\it ad hoc} way  by an excluded volume picture. 
A more systematic treatment can be given in a quantum statistical approach where single-nucleon self-energy shifts 
as well as Pauli blocking are included \cite{RMS}.
 
 Recently, interest arose in a systematic treatment of the nuclear matter EOS \cite{SHT}. 
A combination of NSE with the RMF approach was given in Ref. \cite{Hempel} treating the medium effects for clusters via the excluded volume picture. 
Based on a quantum statistical approach, the nuclear matter EOS including light cluster ($A \le4$) formation was given in Ref.~\cite{Typel}. 
The mass fraction of different clusters has been calculated, and it was shown how clusters are suppressed with increasing density due to Pauli blocking. 
This quantum statistical approach that reproduces NSE and virial expansions in the low-density limit has been applied to supernova explosions \cite{SR} 
as well as to HIC to determine the symmetry energy in the low-density region \cite{Natowitz}.
 
In the quantum statistical approach, the interaction between the different clusters is treated by the method of thermodynamic Green functions \cite{FW}. 
Not only the free nucleons are treated as quasiparticles with momentum dependent energy shifts, 
but also the clusters are described as quasiparticles with energy shifts depending on the center of mass momentum $P$ of the cluster. 
The composition of warm nuclear matter as well as the macroscopic properties are determined by the cluster quasiparticle energy shifts. 
The aim of the present work is to provide the medium modification of the energies of light elements  ($A \le4$) in dense nuclear matter 
in a compact form, that can be easily used in further calculations. 
We generalize and improve previous parametrizations of the quasiparticle energies of light elements \cite{R} as a function of $P, T,n$, and $Y_p$. 
Our goal is to achieve an accuracy of better than 1\% in a wide region of parameter values.

The present work gives the quasiparticle energy shifts only for the light elements. 
Heavier elements need further investigation and are subject to future work. They are of importance at low temperatures. 
Regions of the phase diagram that are dominated by light elements are shown, e.g., in \cite{Hempel}. 
We also neglect weak interaction effects that lead to  $\beta$ equilibrium, as well as Coulomb interaction. 
A homogeneous background of electrons is assumed to compensate the charge of protons. 
The homogeneous nuclear matter considered here may become thermodynamically unstable against phase separation. 
Then, large droplets and more complex structures can be formed, and Coulomb interaction has to be taken into account (see \cite{pasta} and references therein) which, 
however, is not included in our treatment. Also the formation of quantum condensates (pairing, quartetting) will not be discussed here.
 
\section{Quantum statistical approach to warm nuclear matter}

We shortly sketch the quantum statistical approach to give the basic relations, for an extended description see Ref.~\cite{R}.
We consider only strong interaction and neglect $\beta$ equilibrium due to weak interaction. 
In addition to temperature, the total numbers of protons and neutrons as conserved quantities are fixed.  
In the grand canonical ensemble, the corresponding chemical potentials $\mu_n, \mu_p$ are introduced. 
The relation to the nucleon numbers is given by the equation of state \cite{FW}
 \begin{equation}
n_\tau(T, \mu_p, \mu_n)=\frac{2}{\Omega}\sum_{\vec p} \int \frac{d \omega}{2 \pi} f_\tau(\omega) A_\tau(\vec p, \omega)
\end{equation}
where $\Omega$ denotes the system volume. The single nucleon state $\{\vec p, \sigma, \tau\}$ contains the momentum $\vec p$ and 
the isospin $\tau =n,p$. The summation over spin $\sigma$ gives the factor 2. 
The Fermi function reads $f_\tau(\omega)=[\exp((\omega - \mu_\tau)/T)+1]^{-1}$, 
and  $A_\tau(\vec p, \omega)$ is the spectral function of the nucleons. 

The spectral function  $A_\tau(\vec p, \omega)$ is related to the self-energy $\Sigma_\tau(\vec p, \omega)$ according to 
 \begin{equation}
A_\tau(\vec p, \omega)=\frac{2 {\rm Im}\,\Sigma_\tau(\vec p, \omega-i \epsilon)}{[\omega-E_\tau(\vec p)
-{\rm Re}\,\Sigma_\tau(\vec p, \omega)]^2+[{\rm Im}\,\Sigma_\tau(\vec p, \omega-i \epsilon)]^2}\,,
\end{equation}
where $E_\tau(\vec p)= p^2/2m_\tau$ (non-relativistic case). For small ${\rm Im}\,\Sigma_\tau(\vec p, \omega-i \epsilon)$, 
a peak arises at the quasiparticle energy 
$E^{\rm qp}_\tau(\vec p)=E_\tau(\vec p)+{\rm Re} \Sigma_\tau(\vec p, E^{\rm qp}_\tau(\vec p))$. The single-nucleon
quasiparticle energy that depends on the properties of the nuclear medium is usually taken in the effective mass approximation
$E^{\rm qp}_\tau( p; T,n,Y_p)= \Delta E^{\rm SE}_\tau( T,n,Y_p)+p^2/2 m^*_\tau( T,n,Y_p)$ 
where $ \Delta E^{\rm SE}_\tau$ denotes the (rigid) self-energy shift and $m^*_\tau$ the effective mass of the nucleons.

In warm nuclear matter, significant contributions to the spectral function are due to the imaginary part of the self-energy. 
Within a cluster decomposition, see \cite{R}, the $A$-nucleon propagators arise. 
The corresponding $A$-nucleon spectral function describes also the formation of bound states, 
that appear as peaks at  $E_{A,\nu}(\vec P)=E_{A,\nu} + P^2/(2Am)$ in the low-density limit. 
The index $\nu$ denotes the internal quantum state of the $A$-nucleon cluster, $\vec P$ its c.o.m.\ momentum. 
In general, the spin and internal excitation states have to be considered to specify the internal quantum numbers.
This way, the correct low-density limit of warm nuclear matter as a mixture of different components, the nuclei, is obtained.

The $A$-nucleon spectral function is modified with increasing density. 
Similar to the quasiparticle peak in the single-nucleon spectral function, 
the peaks in the  $A$-nucleon spectral function are shifted to quasiparticle energies 
$E^{\rm qp}_{A,\nu}(\vec P)=E_{A,\nu}(\vec P)+{\rm Re}\, \Sigma_{A,\nu}(\vec P, E^{\rm qp}_{A,\nu}(\vec P))$. 
It is clear in other many-particle systems, such as plasmas or condensed matter, 
that bound states behave like new components and can be treated in the same way as elementary particles (chemical picture). 
Notice that the introduction of the quasiparticle concept for the light clusters is related 
to the behavior of the corresponding spectral functions, irrespective of any particular approximation, 
and can be related to measurable properties.

To evaluate the self-energy, perturbation theory and diagram techniques can be used. The cluster expansion of the self-energy \cite{R}
yields the EOS 
\begin{eqnarray}
  n_p(T,\mu_p,\mu_n)&=& {1 \over \Omega} \sum_{A,\nu,P}Z 
f_{A,Z}[E^{\rm qp}_{A,\nu}(P)]\,, \nonumber\\  n_n(T,\mu_p,\mu_n)&=& {1 \over \Omega} \sum_{A,\nu,P}(A-Z) 
f_{A,Z}[E^{\rm qp}_{A,\nu}(P)]\,,
\label{quasigas}
\end{eqnarray}
where
\begin{equation}
\label{faz}
f_{A,Z}(\omega) =
[\exp( \beta (\omega -Z \mu_p -(A-Z) \mu_n) )-(-1)^A]^{-1}
\end{equation}
is the Fermi or Bose distribution function that depends on the inverse temperature $\beta = 1/(k_BT)$ and the chemical
potentials $ \mu_p,  \mu_n$ (instead of the isospin quantum number we use the charge number $Z$). 
The internal quantum number $\nu$ denotes the excited states of the cluster $A,Z$, including the continuum states. 
In addition to the free nucleons $A=1$, where $Z=0$ for $n$ and $Z=1$ for $p$, all higher clusters are included.

We arrive at the NSE in the low-density limit where the quasiparticle energies $E^{\rm qp}_{A,\nu}(\vec P)$ 
can be replaced by the energies $E_{A,\nu}(\vec P)$ of free nucleons and clusters.
The account of the contribution of the continuum (scattering states) in the sum over $\nu$ leads for $A=1,2$ 
to the Beth-Uhlenbeck formula for the second virial coefficient \cite{SRS}.

Quasiparticle shifts have to be considered in warm nuclear matter for densities above $10^{-4}$ fm$^{-3}$. 
Single-nucleon quasiparticle energy shifts $E^{\rm qp}_{\tau}(p)$ have been extensively investigated. 
Various approaches are used such as Skyrme, RMF, or DBHF approximations. These are summarized in Sec.~\ref{sec:introduction} 
and will not be discussed here further. 
We elaborate expressions for the cluster quasiparticle energies $E^{\rm qp}_{A,\nu}(P)$, $A>1$. 
Starting from a Bethe-Salpeter equation for the $A$-nucleon propagator, the treatment of light clusters in warm nuclear matter 
is traced back to the solution of the 
in-medium Schr\"odinger equation 
\begin{equation}
\sum_{1'...A'} H^{\rm matter}(1\dots A;1' \dots A') \psi_{A \nu P}(1'\dots A') =  E^{\rm qp}_{A, \nu}(P) \psi_{A \nu P}(1\dots A)\,,
\end{equation}
where $H^{\rm matter}(1\dots A;1' \dots A')$ is the instantaneous part of the in-medium Hamiltonian. 
In cluster mean-field approximation, explicit expressions are given in Refs.~\cite{R,cmf,DRS}.

\section{Light nuclei in matter: The in-medium effective Schr\"odinger equation  }

The few-body problem describing $A \le 4$ nucleons in hot and dense matter can be related to an in-medium wave equation
(Bethe-Salpeter equation) that is derived from many-particle approaches, see Ref.~\cite{R}.
We consider only bound states and include $A$ in the quantum state $\nu$, dropping the spin quantum number.
For the light elements, $\nu= d, t, h, \alpha$ denotes the deuteron ($^2$H), the triton ($^3$H), the helion ($^3$He), and the $\alpha$ particle  ($^4$He). 
Considering uncorrelated nucleons in the medium, the few-nucleon wave function 
and the corresponding eigenvalues follow from solving the in-medium
Schr\"odinger equation 
\begin{eqnarray}
&&[E_1^{\rm qp}(1)+\dots + E_1^{\rm qp}(A)]\psi_{\nu P}(1\dots A)
\nonumber \\ &&
+\sum_{1'\dots A'}\sum_{i<j}[1- f_1(i)-  f_1(j)]V(ij,i'j')\prod_{k \neq 
  i,j} \delta_{kk'}\psi_{ \nu P}(1'\dots A')
\nonumber \\ &&
= E^{\rm qp}_{\nu}(P) \psi_{ \nu P}(1\dots A)\,.
\label{waveA}
\end{eqnarray}
For  brevity, the single-nucleon quasiparticle energy $ E_{\tau_1}^{\rm qp}(\vec p_1) $ is denoted as $ E_1^{\rm qp}(1) $.
The nucleon-nucleon interaction $V(ij,i'j')$ becomes medium dependent due to the Pauli blocking prefactor 
$[1- f_1(i)- f_1(j)]$.
The phase space occupation is described by a Fermi distribution function normalized to the total density of nucleons,
\begin{equation}
\label{f1}
 f_1(1) = \frac{1}{\exp[E_1^{\rm
qu}(1)/T-  \mu_\tau/T] +1} \approx \frac{n_\tau}{2} \left(\frac{2 \pi \hbar^2}{m^{*}_\tau T}\right)^{3/2} 
e^{-\frac{ p_1^2}{2 m^{*}_\tau T}}
\end{equation}
in the low-density, non-degenerate limit ($ \mu_\tau <0  $). The chemical potential  $ \mu_\tau$ is
determined by the normalization condition $2 \Omega^{-1} \sum_p f_1(p) =
n_{\tau}$, where $\tau $ denotes isospin (neutron or proton), and has to be expressed in terms of these densities and the temperature. 
We restrict our calculations to the low-density region where the non-degenerate limit can be used, 
and replace the effective nucleon masses $m^*_\tau$ by the averaged free nucleon mass $m$ so that $\hbar^2/m = 41.46$ MeV fm$^2$.

The  in-medium Schr\"odinger equation (\ref{waveA}) contains the effects of the medium in the single nucleon quasiparticle shift 
as well as in the Pauli blocking terms. 
Obviously, the bound state wave functions and energy eigenvalues as
well as the scattering phase shifts depend on temperature
and density. In particular, we obtain the cluster quasiparticle shifts 
\begin{equation}
\label{Paul}
E_\nu^{\rm qp}(P;T,n,Y_p)-E_\nu(P)
=\Delta E_\nu^{\rm SE}(P;T,n,Y_p)+\Delta E_\nu^{\rm Pauli}(P;T,n,Y_p). 
\end{equation}
Of special interest are the binding energies $ E^{\rm bind}_{\nu}(P)= E^{\rm qp}_{\nu}(P)-A E_1^{\rm qp}(P/A)$ 
that indicate the energy difference between the bound state and the continuum of free (scattering) states at the same total momentum $P$. 
This binding energy determines the yield of the different nuclei according to Eq.~(\ref{quasigas}), where the summation over $P$ is restricted to that region where bound states exist, i.e. $E^{\rm bind}_{\nu}(P) \le 0$. 

The contribution of the single nucleon  energy shift to the cluster self-energy shift $\Delta E_\nu^{\rm SE}$ 
is easily calculated in the effective mass approximation, 
where the single-nucleon quasiparticle energy shift $\Delta E_1^{\rm SE}(1)$ 
can be represented by the energy shift 
$\Delta E_\tau^{\rm SE}$ and the effective mass $[m^*_\tau]^{-1}=[m_\tau]^{-1}+\partial^2 \Delta E_\tau^{\rm qp}(p)/ \partial p^2|_{p=0}$.  
In the rigid shift approximation where $m^*=m$, the energy shift 
$\Delta E_\tau^{\rm SE}$ cancels in the binding energy and can be absorbed in the chemical potential of the EOS (\ref{quasigas}). 
The effective mass correction is also easily calculated for given wave functions, see \cite{SR,R}. 
The influence of the self-energy shifts on the binding energy is small and will be considered below in Sec. \ref{SEsec}.

We consider here the Pauli blocking shift of the binding energies 
$\Delta E_\nu^{\rm Pauli}(P;T,n,Y_p)=E_\nu^{\rm qp}(P)-E_\nu(P)- (A-Z) \Delta E_n^{\rm SE}- Z \Delta E_p^{\rm SE}\,.
$
To evaluate it, we need the interaction potential $V(ij,i'j')$. We choose a simple separable interaction potential with Gaussian form-factors
\begin{eqnarray}
&& V(p_i,p_j;p_i',p_j') = \lambda_\nu \delta_{p_i+p_j,p_i'+p_j'} 
e^{-\frac{(p_i-p_j)^2}{4 \gamma_\nu^2}} e^{-\frac{(p_i'-p_j')^2}{4 \gamma_\nu^2}} \,.
\end{eqnarray}
The two parameters $\lambda_\nu$ and $\gamma_\nu$ can be fitted to reproduce the binding energy and the root mean square (rms) radii 
of the free nuclei as empirical input \cite{atomdata,Wiringa}.

The solution of the few nucleon problem in the low-density limit can be found from  variational,   Faddeev  \cite{Beyer,Sedrakian},  
Green's-function Monte Carlo \cite{Wiringa}, etc., approaches. 
We use a variational approach with the Jastrow ansatz \cite{R} for $\psi_{ \nu P}(1\dots A)$ that reproduces the exact solution for $A=2$,
\begin{equation}
\label{varphi}
\varphi^{\rm Jastrow}_\nu(\vec p_1\dots \vec p_A) = \frac{1}{N_\nu}\prod_{i<j}
\frac{e^{-\frac{(\vec p_j-\vec p_i)^2}{4a_\nu^2}}}{\frac{(\vec p_j-\vec p_i)^2}{4 b_\nu^2 }+1}\,.
\end{equation}
The prefactor $N_\nu$ is determined by normalization.

Details are given in Ref. \cite{R} where Jacobian coordinates are introduced. 
The fit of the potential and the corresponding wave functions are given in Tab.~\ref{Tab1}. 
Due to the tensor force, the nucleon-nucleon interaction in the spin-triplet state is stronger than in the spin-singlet state. 
As detailed in \ref{app1},
different orbitals for protons and neutrons will occur for $A=3$, and different rms radii for protons and neutrons are considered there. The present analysis is based on  averaged rms radii 
for $t,h$ given in Tab.~\ref{Tab1}.
\begin{table}
\caption{ Light cluster wave function parameter values at zero density related to the Jastrow approach Eq.~(\ref{varphi})}
\begin{center}
\hspace{0.5cm}
\begin{tabular}{|c|c|c|c|c|c|c|c|}
\hline
 $\nu$ & $\lambda_\nu$ & $\gamma_\nu$ &$ a_\nu$ & $b_\nu$ & $E_\nu^{\rm bind}$ & $E_\nu^{\rm kin}$ & rms$_\nu$ point 
  \\
&[MeV fm$^{3}$] & [fm$^{-1}$] & [fm$^{-1}$] & [fm$^{-1}$] &  [MeV] &  [MeV] &  [fm]  \\
\hline
$d$ ($^2$H) &-1287.4 & 1.474  &  1.474 &  0.2317 &-2.225 & 10.338 & 1.96 \\
$t$ ($^3$H)& -1467.0  & 1.153  & 1.595 & 0.567 & -8.482 & 23.735 & 1.68  \\
$h$ ($^3$He) & -1431.9  & 1.153  & 1.602 & 0.5514 & -7.718 & 23.021 & 1.71   \\
$\alpha$ ($^4$He) & -1272.9  & 1.231  & 2.151 & 0.912 & -28.30 & 51.575 & 1.45   \\
\hline
\end{tabular}
\end{center}
\label{Tab1}
\end{table}

\section{Binding energies at finite densities}

\subsection{Pauli shift in the low-density limit}

We focus on the Pauli blocking shift $\Delta E_\nu^{\rm Pauli}(P;T,n,Y_p)$ that is responsible for the disappearence of bound states. It is calculated from 
Eq.~(\ref{waveA}) when the influence of the medium is accounted for only in the Pauli blocking terms.

The solution is found in the low-density limit by perturbation theory, at arbitrary density by variational calculations. 
We use the Jastrow ansatz (\ref{varphi}) and  determine optimal values for $a_\nu$ and $b_\nu$. 
The evaluation of normalization, kinetic and potential energy is performed using Jacobi coordinates \cite{R}. 
Note that the in-medium Hamiltonian (\ref{waveA}) is not hermitian but can be transformed to a hermitian Hamiltonian,
see \ref{app3}.

The evaluation of the Pauli blocking shift $\Delta E_\nu^{\rm Pauli}(P;T,n,Y_p)$ is rather involved and time consuming. 
In the variational approach, we have to calculate multiple integrals and to search for a minimum. 
Similar to the single nucleon quasiparticle shift, where instead of the more fundamental DBHF calculations a simple
fit within the RMF approach is convenient for further calculations, we will give simple expressions for the Pauli blocking shift
that reproduce the numerical evaluations. 

In the low-density limit, the perturbation in Eq.~(\ref{waveA}) is proportional to the neutron density $(1-Y_p) n$ and the proton density $Y_p n$. We discuss first the dependence on the asymmetry $Y_p$. For the deuteron, Pauli blocking is determined by the sum of the 
neutron and proton distribution functions $f_n(p_1)+f_p(p_2)$, see Eq.~(\ref{waveA}). 
It depends only on the total baryon density $n$ in the non-degenerate limit.
Therefore, we neglect the dependence on $Y_p$ for $\nu =d$. The same applies for $^4$He because neutron and proton orbitals are 
equally occupied for $\nu = \alpha$. In the clusters with $A=3$, however, neutrons and protons contribute differently to the internal structure, 
so that the shifts of $^3$H and $^3$He are sensitive to the asymmetry of nuclear matter.  We find
\begin{eqnarray}
\label{thsym}
\Delta E_\nu^{\rm Pauli}(P;n,T,Y_p)&=& f_{\nu}(P;T,0)\,\, y_\nu(Y_p)\,\, n+ {\cal O}(n^2)\,, 
\end{eqnarray}
with $y_d(Y_p)=y_\alpha(Y_p)=1$, for triton $y_t(Y_p)=\left( \frac{4}{3}-\frac{2}{3} Y_p\right)$, and for helion $y_h(Y_p)=\left(\frac{2}{3}+\frac{2}{3} Y_p\right)$.
For example, in comparison with helions ($^3$He), the tritons ($^3$H) show a stronger 
shift in neutron-rich matter because the neutrons in the cluster are stronger blocked than the protons. 

The functions $f_\nu(P;T,0)$ can be calculated in first order perturbation theory using the unperturbed wave 
functions of the free nuclei. Motivated by the 
exact solution for $A=2$, we use the following fit for arbitrary $\nu$:
\begin{eqnarray}
\label{perturb}
&& f_\nu(P;T,n)= f_{\nu,1}  \exp\left[-\frac{P^2/\hbar^2}{4 ( f_{\nu,4}^2/f_{\nu,3}^2) (1+T/f_{\nu,2})+u_\nu n}\right] \frac{1}{T^{1/2}} \frac{2 f_{\nu,4}}{P/\hbar}   \nonumber\\
&&\times {\rm Im} \left\{ \exp\left[f_{\nu,3}^2 (1+f_{\nu,2}/T) \left(1-i\frac{P/\hbar}{2 f_{\nu,4}(1+T/f_{\nu,2})}\right)^2\right] \right. \nonumber\\
&&\left. \times {\rm erfc}\left[ f_{\nu,3} (1+f_{\nu,2}/T)^{1/2} \left(1-i\frac{P/\hbar}{2 f_{\nu,4}(1+T/f_{\nu,2})}\right)\right]\right\}
\end{eqnarray}
where the term $u_\nu n$ can be neglected in the low-density limit considered in this subsection.
\begin{table}
\caption{ Parameter values for the Pauli blocking shift $\Delta E^{\rm Pauli}_\nu(P;T,n,Y_p)$, Eq.~(\ref{perturb}), in the low-density limit}
\begin{center}
\hspace{0.5cm}
\begin{tabular}{|c|c|c|c|c|c|}
\hline
 $\nu$  & $f_{\nu,0}$  &$ f_{\nu,1}$  & $f_{\nu,2}$ & $f_{\nu,3}$ & $f_{\nu,4}$ \\
 & [MeV fm$^{5/2}$]& [MeV fm$^3$] & [MeV] & -& [fm$^{-1}]$\\
\hline
$d$ ($^2$H) & 388338  & 6792.6 &  22.52 &0.2223 &0.2317\\ 
 $t$ ($^3$H) &  159080 &   20103.4 & 11.987 &0.85465  &0.9772\\
 $h$ ($^3$He) &153051 &19505.9 &  11.748  & 0.84473  &0.9566 \\
$\alpha$ ($^4$He)  & 352965 & 36146.7 & 17.074  &0.9865&1.9021\\
\hline
\end{tabular}
\label{Tab.2neu}
\end{center}
\end{table}

In particular, at zero c.o.m. momentum $P=0$ we have
\begin{equation}
\label{ffit}
f_\nu(0;T,0)= \frac{f_{\nu,0}}{(f_{\nu,2}+T)^{3/2}}F(x_\nu), \qquad x_\nu=f_{\nu,3}(1+f_{\nu,2}/T)^{1/2}\,,
\end{equation}
where $F(x)=2 x^2 \left(1-\pi^{1/2} x e^{x^2} {\rm erfc}(x) \right)$.
The complementary error function is defined as  $ {\rm erfc}(x) =1-{\rm erf}(x) = 1-2 \pi^{-1/2}\int_0^x \exp(-t^2) {\rm d} t$
so that $\lim_{x \to \infty}F(x)=1-3/(2 x^2) \pm \dots$, 
$\lim_{x \to 0}F(x)=2 x^2 (1- \pi^{1/2} x +2 x^2 - \pi^{1/2} x^3 \pm \dots)$. 
Expression (\ref{ffit}) is exact for the two-nucleon case, where (in units of MeV, fm) $a_d=\gamma_d=1.474,
\,\,\,b_d=f_{d,4}=(-E_d^{\rm bind} m/\hbar^2)^{1/2}
=0.2317,\,\,\,f_{d,2}=\hbar^2 a_d^2/(4m) = 22.52,
 \,\,\, f_{d,3} = 2^{1/2} b_d/a_d = 0.2223, $ 
$$
f_{d,0}=\left(\frac{\hbar^2}{m}\right)^{5/2} \frac{2^{1/2} \pi^{3/2} a_d^2}{2-(2+1/f_{d,3}^2) F(f_{d,3})}=388338\,,
$$
and
$$
f_{d,1}= \left(\frac{\hbar^2}{m}\right)^{3/2} \frac{ \pi^2 2^{3/2} f_{d,3} }{1-(1+1/(2 f_{d,3}^2)) F(f_{d,3})} 
=6792.6\,.
$$
The corresponding parameter values for the other light nuclei are given in Tab.~\ref{Tab.2neu}. 
The intended accuracy to reproduce the solution of Eq.~(\ref{waveA}) is better than1\% in the parameter region considered here.

Note that the dependence of the Pauli blocking shifts on the asymmetry parameter $Y_p$ becomes more involved if we consider different orbitals for the neutrons and the protons in the clusters. In particular, this applies
for triton ($^3$H) and helion ($^3$He) where the neutron and proton orbitals are different, see \ref{app1}.

\subsection{Pauli shift at arbitrary subsaturation density}

The results for the Pauli shift obtained in the low density limit using perturbation theory can be extended to higher densities using a variational approach to solve the in-medium Schr\"odinger equation Eq.~(\ref{waveA}). We used the Jastrow ansatz (\ref{varphi}) for the class of wave functions and optimized the parameter values for $a_\nu, b_\nu$.

In the low-density limit, a linear dependence of the energy shifts 
on the nucleon density $n$ follows from  perturbation theory. For a more general dependence on the total nucleon density
we consider the expression
\begin{eqnarray}
\label{delpauli0P2}
&&\Delta E_\nu^{\rm Pauli}(P;n,T,Y_p)= 
c_{\nu}(P;T) \left\{1-\exp\left[- \frac{f_\nu(P;T,n)}{c_{\nu}(P;T)} y_\nu(Y_p)n- d_{\nu}(P;T,n)n^2 \right] \right\}\,.
\end{eqnarray}
This ansatz has been taken to ensure that the shift will not decrease with increasing density 
and to avoid spurious reappearance of bound states 
at high densities, after they are blocked out. The linear term $f_\nu(P;T,0)$ is given by first order perturbation theory with respect to the density.
The asymptotic term $c_{\nu}(P;T)$ should be larger than the cluster binding energy so that the bound state merges with
the continuum of scattering states at a certain density and disappears. We restrict ourselves to the quadratic term with respect to the density dependence, a further density dependence in the prefactor $d_{\nu}(P;T,n)$ disappears at $P=0$ as discussed below.

We have reduced the parameter dependences as much as possible to get the intended accuracy of few percent 
to reproduce the Pauli shifts also at higher densities. For zero momenta, $P=0$, the temperature dependence of $c_{\nu}(0;T)$ and $d_{\nu}(0;T,n)$ is expressed as

\begin{equation}
c_{\nu}(0;T)=c_{\nu,0}+\frac{c_{\nu,1}}{(T-c_{\nu,2})^2+c_{\nu,3}},\qquad d_{\nu}(0;T,n)=\frac{d_{\nu,1}}{(T-d_{\nu,2})^2+d_{\nu,3}}\,.
\end{equation}
The corresponding parameters are given in Tab.~\ref{Tabcd}.

The Pauli shift at finite momenta is fitted with  $c_{\nu}(P;T)=c_{\nu}(0;T)$ not depending on $P$, but
\begin{equation}
 d_{\nu}(P;T,n)=d_{\nu}(0;T,n)e^{-\frac{P^2/\hbar^2}{v_\nu T n}}\,.
\end{equation}
An additional dependence on $n,T$ is considered at finite values of $P$. 
Another additional dependence on $n,T$ for finite momenta is introduced in $f_\nu(P;T,n)$ where the dispersion relation becomes density dependent due to the parameter $u_\nu$.
Parameter values are given in Tab.~\ref{Tabcd}. Starting from the fit to deuterons, the parameter values $u_\nu$ describing the dispersion were scaled to $A^2$, see also \cite{Typel} where similar values have been considered. The optimalisation of the parameter values $v_\nu$ was performed to reproduce the calculated values for the Pauli blocking shift within few percent. 

\begin{table}
\caption{ Parameter values for the Pauli blocking shift $\Delta E^{\rm Pauli}_\nu(0;T,n,Y_p)$, Eq.~(\ref{delpauli0P2}),
in units of MeV and fm ($c_{\nu,0},c_{\nu,2},d_{\nu,2}$ - [MeV];
$c_{\nu,1}$ - [MeV$^3$]; $c_{\nu,3},d_{\nu,3}$ - [MeV$^2$]; $d_{\nu,1}$ - [MeV$^2$ fm$^6$]; $u_\nu$ - fm; $v_\nu$ - [MeV$^{-1}$ fm])}
\begin{center}
\hspace{0.5cm}
\begin{tabular}{|c|c|c|c|c|}
\hline
 $\nu$  &$d$ ($^2$H) & $t$ ($^3$H) & $h$ ($^3$He)&$\alpha$ ($^4$He) \\
\hline
 $c_{\nu,0}$ &2.752 &11.556&10.435 &150.71 \\
 $c_{\nu,1}$ &32.032  & 117.24 &176.78  &9772 \\
 $c_{\nu,2}$   & 0 & 3.7362 &  3.5926 &2.0495 \\
$c_{\nu,3}$ & 9.733 & 4.8426 & 5.8137  &2.1624\\
\hline
$d_{\nu,1}$  & 523757  & 108762 & 90996  &5391.2 \\
 $d_{\nu,2}$  &0&  9.3312 &  10.72 &    3.5099 \\
 $d_{\nu,3}$ & 15.273 &49.678 &47.919 & 44.126\\
\hline
$u_{\nu}$  &11.23  &25.27 & 25.27  &44.92 \\
\hline
$v_{\nu}$  &0.145  &0.284 & 0.27  &0.433 \\
\hline
\end{tabular}
\label{Tabcd}
\end{center}
\end{table}

\subsection{\label{SEsec} Self-energy shift}

The quasiparticle shifts, Eq.~(\ref{Paul}), contain besides the Pauli blocking also self-energy terms.
The self-energy shifts can be easily evaluated if the effective mass approximation is taken for the single nucleon quasiparticle dispersion relation.  Taking the unperturbed Jastrow wave function (\ref{varphi}),
the perturbative treatment gives the result
\begin{equation}
\Delta E_\nu^{\rm SE}(P;T,n,Y_p)=\left(E_\nu^{\rm kin}+\frac{P^2}{2 A_\nu^2 m} \right) \left( \frac{m}{m^*}-1 \right)
\end{equation}
where the kinetic energy $E_\nu^{\rm kin}$ of the internal motion of the cluster $\nu$ is given in  Tab.~\ref{Tab1}.
It results as the averages of $\hbar^2/m q_1^2$ for $A=2$, $\hbar^2/m (q_1^2+3/4 q_2^2)$ for $A=3$, 
and $\hbar^2/m (q_1^2+3/4 q_2^3 +2/3 q_3^2)$ for $A=4$,
where $\vec q_i$ denote the respective Jacobian momenta.
To estimate the effect, 
an empirical expression  $m^*/m=1-0.17 n/n_{\rm sat}$ can be taken.

We denote the momentum $P_d^{\rm Mott}(T,n,Y_p)$, where the bound state disappears, as Mott momentum, see \ref{app2}. At $n>n_d^{\rm Mott}(T,Y_p)$,
the summation over the momentum to calculate the bound state contribution to the composition is restricted to the region 
$P >P_d^{\rm Mott}(T,n,Y_p)$. Below the Mott density, the influence of the self-energy shifts is small. If the nucleon density approaches the saturation density, the suppression of clusters is enhanced due to the self-energy shifts.

\section{Discussion and outlook}

The context of this work is the calculation of the composition of 
warm nuclear matter covering a wide range of parameter values $T, n$, and $ Y_p$.
Within a quantum statistical approach, exact results in the low-density limit 
as well as quasiparticle approaches near the saturation density are 
reproduced. A generalized Beth-Uhlenbeck approach \cite{SRS} can be used that
leads to NSE and the virial EOS at low densities, but includes medium effects
that become relevant for baryon number densities above $\sim 10^{-4}$ fm$^{-3}$.

In the present work, we focus on the light elements $d,t,h,\alpha$ immersed in warm and dense matter.
The quasiparticle description is an important prerequisite for their proper treatment
in connection with thermodynamic or transport properties. The parametrization 
given here can be improved in different ways. The solution of the few-body problem 
accounting for medium effects, Eq.~(\ref{waveA}), can be performed using more sophisticated 
methods instead of the variational approach. The class of functions that represent the quasiparticle
shifts as function of temperature, density, and asymmetry parameter, 
can be extended and better adjusted to the solution of the
few-body problem for the light elements.

We restrict us to the light elements $A\le4$. An extension to heavier 
elements is possible but needs further considerations, see Ref. \cite{roe84}.
Since the following elements up to carbon are weakly bound, only small yields 
are expected.
In particular, at low temperatures the formation of  heavier clusters is of interest.
Then, the proper treatment of Coulomb interactions and inhomogeneous 
solutions is inevitable \cite{LS,Shen,SHT}. As well known, in the region of thermodynamical
instability the Coulomb interaction is responsible for structure formation representing 
both phases.

Another issue is the treatment of excited states, in particular the continuum of scattering states,
that are also included in the EOS (\ref{quasigas}). The contribution of scattering states 
has to be taken into account to reproduce the correct low-density limit of the
second virial coefficient. For the in-medium
treatment of scattering states in the two-nucleon case, as well as the
evaluation of the second virial coefficient, see \cite{RMS,SRS,HS}. In particular,
the deuteron fraction is reduced if scattering states are considered, 
because the binding energy is comparable to the temperature. Sharply peaked structures
related to the Mott effect are washed out because of the Levinson theorem, see \cite{SRS}.
For nonequilibrium processes like HIC, the pole structure of the cluster spectral function may be relevant so that 
the cluster yields are not determined by  the contribution of scattering states to the second virial coefficient.

The extension of the low-density results for the quasiparticle shifts to the region of the 
saturation density needs further discussion. The expressions for the self-energy and the Pauli blocking
are obtained in first order with respect to the distribution function $f_1(1)$ in the medium part of the 
Hamiltonian. Higher orders will 
arise, e.g., from the free $A$-particle propagator. Also, the peaks in the spectral function that 
characterize the quasiparticle excitations are broadened at increasing density what reflects the
damping of the quasiparticles due to collisions.

Moreover, the approximation of the uncorrelated medium used in the in-medium 
wave equation (\ref{waveA}) can be improved 
considering the cluster mean-field approximation \cite{RMS,R,cmf}. 
In particular, at low temperatures and low densities $\alpha$ cluster become dominant,
and the medium effects of $\alpha$ matter are produced by the surrounding $\alpha$ clusters. 
Then, the $\alpha$ quasiparticle energy shift is determined by the effective $\alpha-\alpha$ interaction.
Finally, degeneracy and the formation of quantum condensates (pairing, quartetting) 
has to be considered in the low temperature region, when the single-nucleon chemical potentials 
approach the lowest bosonic ($A=2,4$) binding energy. 

In conclusion, the parametrization of the quasiparticle cluster energies provides us with 
a tool to extend low-density approaches such as NSE to higher densities where
mean-field effects become relevant. This is an indispensable ingredient to calculate the 
composition of nuclear matter in astrophysical applications as well as in HIC. 
The contributions to the quasiparticle energies that are linear in the density, expressed by the functions
$f_\nu(P;T)$, Eq.~(\ref{delpauli0P2}), are well described already in 
perturbation theory. At higher densities, exploratory results have been given, in particular with
respect to the disappearance of bound states with given c.o.m.\ momentum. The detailed treatment of correlations
formed in the nucleonic many-body system at higher densities, 
in particular the evaluation of the spectral function, remains a challenging problem.


\appendix

\section{\label{app1} Binding energies and rms radii of light nuclei}

We start out from the empirical values for the binding energies and charge rms radii for the light elements \cite{atomdata,Wiringa}, see Table~\ref{emp}. 
The experimental charge radii have been converted to point proton rms radii by removing the proton and neutron $\langle r^2 \rangle$
of 0.743 and -0.116 fm$^2$, respectively.

Because of equal numbers, the neutron and proton orbitals are approximately identical in $d$ and $\alpha$. 
The situation is more complex for $t$ and $h$ where the numbers of protons and neutrons are different. 
We give some exploratory calculations to estimate the different orbitals of neutrons and protons in $t$ and $h$, 
in particular the different rms radii.

\begin{table}
\begin{center}
\caption{Empirical data of light nuclei \cite{atomdata,Wiringa}}
\vspace{0.5 cm}
\begin{tabular}{|l|l|l|c|r|r|}
\hline
  & binding & mass & spin & rms-radius & rms-radius
  \\
  & energy&  &  &  (charge)& (point, proton)\\ 
&[MeV] &[MeV/$c^2$] & & [fm] &[fm]  \\
\hline
$n$ & 0 & 939.565  & 1/2&0.34  & 0  \\
$p$ ($^1$H) & 0 & 938.783  & 1/2 & 0.87  & 0 \\
$d$ ($^2$H) & -2.225 & 1876.12  & 1 & 2.12  & 1.96 \\
$t$ ($^3$H) & -8.482  & 2809.43  & 1/2 & 1.76  & 1.58   \\
$h$ ($^3$He) & -7.718  & 2809.41  & 1/2 & 1.93  & 1.76   \\
$\alpha$ ($^4$He) & -28.30  & 3728.40  & 0 & 1.68   & 1.48  \\
\hline
\end{tabular}
\label{emp}
\end{center}
\end{table}


We deduce characteristics of the nucleon wave functions from experimental data that are available for 
the binding energies 
$E_t, E_h$, as well as the proton rms radii ${\rm rms}_{t,p}$ and $ {\rm rms}_{h,p}$.
Coulomb interactions and tensor forces are responsible for the differences between $t$ and $h$. 
As mirror nuclei, the difference in the binding energy is caused by the Coulomb interaction that leads 
to a repulsion between the protons in $^3$He so that the binding energy is reduced if comparing with $^3$H. 
We also expect that the rms values are slightly larger in $h$ because it is not so strongly bound as $t$. 
However, the observed large difference in the rms radii is due to the tensor force of the nucleon-nucleon 
interaction that is related to spin orientation.

The occurrence of the deuteron as a bound state in the $p-n$ spin triplet channel, 
in contrast to $p-p$ or $n-n$ where a bound state is absent, 
is due to the tensor force that makes the interaction between nucleons in the spin-triplet state stronger 
than between nucleons in the spin-singlet state. 
In the ground state of $^3$H, there are two neutrons with opposite spin orientation, 
whereas a proton is found, say, with spin up. The interaction of this proton with the neutrons 
contains contributions from the tensor force (with the spin-up neutron). 
The spin-up neutron feels also the tensor force (with the proton), 
whereas the spin-down neutron sees only opposite spins so that no tensor force acts. 
On average, the neutron is weaker bound in $^3$H than the proton so that its rms radius is larger,  
${\rm rms}_{t,n} > {\rm rms}_{t,p}$. 
Vice versa, in $^3$He the proton rms radius is expected to be larger than the neutron one, 
${\rm rms}_{h,p} > {\rm rms}_{h,n}$. 

To separate both, the tensor and Coulomb effects, we assume first that Coulomb effects can be discarded. 
Then we have ${\rm rms'}_{t,n} = {\rm rms}_{h,p}$ and  ${\rm rms'}_{h,n} = {\rm rms}_{t,p}$.  
In the average, the point nucleon rms radius of the three-nucleon bound state is 
${\rm rms'}_{3,{\rm aver}} = ({\rm rms}_{t,p} + 2\,\, {\rm rms}_{h,p})/3 =$ 1.70 fm. 
Correspondingly, we consider an averaged binding energy $E_{3,{\rm aver}} = (E_t + 2\, E_h)/3 =$ 7.973 MeV. 
This averaged three-nucleon bound state serves only for guidance to estimate the effects and does not provide high precision, 
because it is contaminated by Coulomb effects. 
For this average nucleus, we obtain the parametrization $\lambda_3=1443.66, \gamma_3=1.153, a_3= 1.5998$, and $ b_3=0.5567$.

For the further analysis, we assume that the range of the interaction $\gamma$ is nearly constant. 
To study the influence of the Coulomb interactions we change the strength $\lambda$ of the interaction 
and evaluate the change of binding energy and of the rms radius. 
This gives a relation between the binding energy and the rms radius. 
In particular, for $E_3=E_t$ we find ${\rm rms}_{t}$  = 1.68 fm, 
and for  $E_3=E_h$ we find ${\rm rms}_{h} =$ 1.71 fm. 
From this, we conclude that the weaker bound nucleus ($h$) is blown up by ${\rm rms}_{h} -{\rm rms}_{t}$ = 0.03 fm. 
We determine the potential parameters $\lambda, \gamma$ that reproduce the values 
$E_t, {\rm rms}_{t}$, and $E_h, {\rm rms}_{h}$, see Table~\ref{Tab1}. 

Note that we can specify the neutron point rms radius, 
assuming that the orbits in $h$ are expanded by 0.03 fm compared to $t$. 
Using the measured rms radii for the protons, 
we estimate values for the neutron point rms radius ${\rm rms}_{t,n}, {\rm rms}_{h,n}$ shown in the Table~\ref{Tab1a}. 
Finally, we determined the proton and neutron wave functions assuming that the range of interaction 
remains fixed and the strength of interaction is fitted to the respective rms radii. Results are given in Table~\ref{Tab1a}.
Using the corresponding parameter values, we can calculate the contributions of the different orbitals 
to the Pauli blocking shift of the triton $t$ and helion $h$. 
We will give the results in a forthcoming work and restrict us here to the 
shifts derived from the averaged values given in  Table~\ref{Tab1}. 


\begin{table}
\caption{Light cluster wave function parameter at zero density from the Jastrow approach, $A$ = 3}
\begin{center}
\begin{tabular}{|l|l|l|c|l|l|l|}
\hline
 $\nu$ & $\lambda_\nu$ & $\gamma_\nu$ &$ a_\nu$ & $b_\nu$ & $E_\nu$ & rms$_\nu$ point 
  \\
&[MeV fm$^{3}$] & [fm$^{-1}$] & [fm$^{-1}$] & [fm$^{-1}$] &  [MeV] &  [fm]  \\
\hline
$t$ ($^3$H)$p$ & -1600.5  & 1.153  & 1.571 & 0.6237 & -11.525 & 1.59   \\
$t$ ($^3$H)$n$ & -1413.2  & 1.153  & 1.607 & 0.5429 & -7.320 & 1.73   \\
$h$ ($^3$He)$p$ & -1381.5  & 1.153  & 1.615 & 0.5281 & -6.645 & 1.76   \\
$h$ ($^3$He)$n$ & -1553.6  & 1.153  & 1.578 & 0.6041 & -10.441 & 1.62   \\
\hline
\end{tabular}
\label{Tab1a}
\end{center}
\end{table}

\section{\label{app2} Pauli blocking shift of the deuteron}

In the case $A=2$, we introduce Jacobian wave numbers $\vec q_1=(\vec p_2-\vec p_1)/2 \hbar, \hbar \vec q_2 = \vec P=\vec p_1+\vec p_2$.
The exact solution of Eq.~(\ref{waveA}) gives the momentum dependent in-medium binding energy 
$E^{\rm bind}_d(P)=E^{\rm qp}_d(P)-P^2/(4m)=E_d+\Delta E_d^{\rm Pauli}(P)$ according to ($K=P/\hbar$)
\begin{eqnarray}
\label{deu}
&&\frac{\lambda_d}{4 \pi^2} \int_0^\infty dq_1 \int_{-1}^1 dz_1 \frac{q_1^2}{E^{\rm bind}_d(P)-\hbar^2q_1^2/m} e^{-\frac{2 }{ \gamma_d^2} q_1^2}
\nonumber \\
&& \!\!\!\!\!\!\! \!\!\!\!\!\!\!\!\!\!\!\!\!\!\times \left[ 1- \frac{1}{ e^{\frac{\hbar^2}{2 m T} (q_1^2-q_1K z_1+K^2/4)-\frac{\mu_n}{T}}+1 }- 
 \frac{1}{ e^{\frac{\hbar^2}{2 m T} (q_1^2+q_1 K z_1+K^2/4)-\frac{\mu_p}{T}}+1} \right]=1\nonumber \\
&& {}
\end{eqnarray}
with effective chemical potentials (containing the self-energy shifts) determined by
\begin{equation}
 \frac{1}{\pi^2} \int_0^\infty dk k^2 \frac{1}{ e^{\frac{\hbar^2 k^2}{2 m T} -\frac{\mu_\tau}{T}}+1 }=n_\tau\,.
\end{equation}
Numerical solution gives the binding energy $E_d^{\rm bind}(P;T,n,Y_p)$.

\begin{figure}
\begin{center}
\includegraphics[width=10cm]{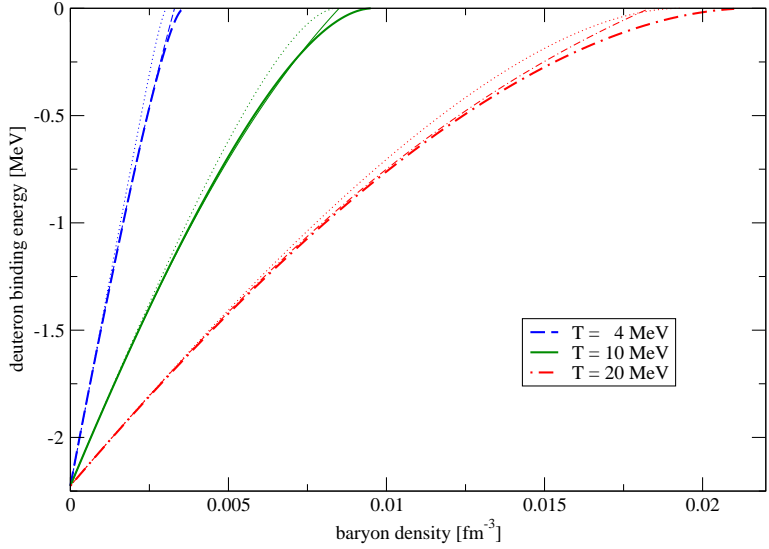}
\caption{\label{fig:1} 
Shift of the deuteron binding energy $ E^{\rm bind}_d(P=0)=E_d +\Delta E^{\rm Pauli}_d(0) $, Eq.~(\ref{Paul}), 
as function of the baryon density $n$ 
at different temperatures $T, Y_p=0.5$. The numerical solution of Eq.~(\ref{deu}) is compared with the fit formula (thin lines), 
Eq.~(\ref{delpauli0P2}), and the non-degenerate approximation (dotted lines).}
\end{center}
\end{figure}

\begin{figure}
\begin{center}
\includegraphics[width=10cm,angle=-90]{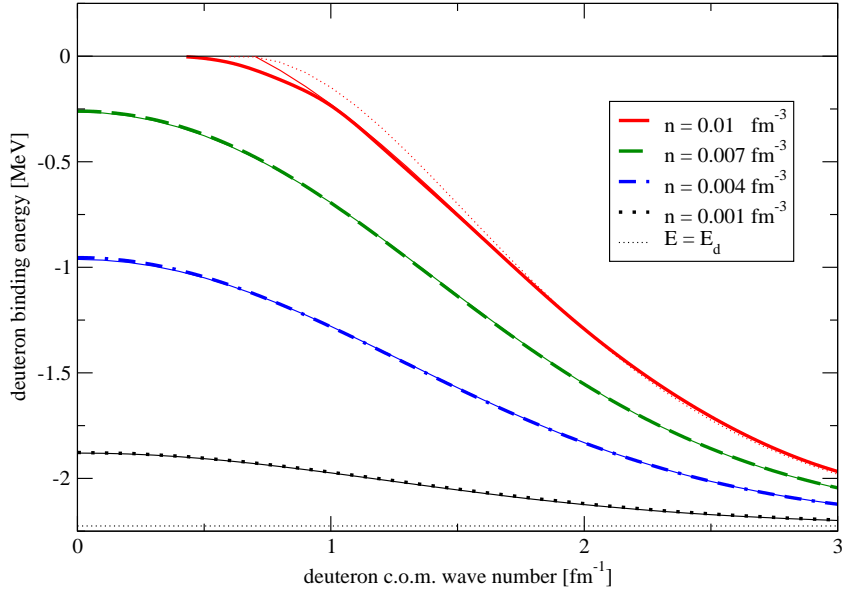}
\caption{\label{fig:2} 
Shift of the deuteron binding energy  $ E^{\rm bind}_d(\hbar K)=E_d +\Delta E^{\rm Pauli}_d(\hbar K) $, Eq.~(\ref{Paul}), 
as function of the deuteron c.o.m.\ momentum $P= \hbar K$ at $T=10$ MeV for different baryon densities $n, Y_p=0.5$. 
 The numerical solution of Eq.~(\ref{deu}) is compared with the fit formula (thin lines), Eq.~(\ref{delpauli0P2}). The dotted line is the fit according Eq.~(\ref{B7}).}
\end{center}
\end{figure}

In the non-degenerate case, {\it i.e.}\ at low densities and not too low temperatures, 
where the effective chemical potentials are negative, integrations can be performed analytically. 
Then, it is easily shown that only the total baryon density $n=n_n+n_p$ enters so that the solution does not depend on the asymmetry $Y_p$. 
The typical behavior of the binding energy is shown for symmetric matter ($Y_p=0.5$) in Figs.~\ref{fig:1} and \ref{fig:2}. 
For comparison, the solution replacing the Fermi function by the Boltzmann function is also shown. At the temperatures considered, the differences are small. 
Only at very low temperatures where Bose-Einstein condensation is possible, significant differences are expected.

Depending on temperature, the absolute value of the binding energy at zero momentum $E_d^{\rm bind}(0;T,n,Y_p)$ is decreasing with increasing density 
and merges with the continuum of scattering states. The analytical fit reproduces this behavior with exception of the region where 
the bound state merges with the continuum. We denote this density as Mott density $n_d^{\rm Mott}(T,Y_p)$. 
Note that this disappearance of a bound state does not introduce discontinuities in the EOS, 
provided that the contribution of scattering phase shifts is also taken into account \cite{SRS}.

To give a simple approximation for the Mott density we perform a fit (in units MeV, fm$^{-3}$), 
\begin{equation}
 n_d^{\rm Mott}(T,Y_p) \approx 0.00057646 + 0.00065443 \,\,T +0.000012491\,\, T^2\,.
\end{equation} 
For baryon densities higher than the Mott density, bound states can be formed only for c.o.m.\ momenta $P$ larger than the 
Mott momentum $\hbar K_d^{\rm Mott}(T,n,Y_p)$. Using the solution of Eq. (\ref{deu}) at arbitrary degeneracy, 
a fit is given as 
\begin{eqnarray}
 &&\left[ K_d^{\rm Mott}(T,n,Y_p)\right]^2 \approx -\frac{4.5185-0.16164\, T+0.0056582\, T^2}{2(1.32-0.02782 \,T)}  \nonumber \\ && \!\!\!\! +
\left(\frac{(4.5185-0.16164\, T+0.0056582\, T^2)^2}{4(1.32-0.02782\, T)^2}+\frac{1000 (n-n_d^{\rm Mott}(T,Y_p))}{1.32-0.02782 \,T}
\right)^{1/2}\,.  \nonumber \\ && {}              
\end{eqnarray} 

The behavior of $E_d^{\rm bind}(P;T,n,Y_p)$ as function of the c.o.m.\ momentum $\vec P$ is shown in Fig.~\ref{fig:2} 
for different densities at fixed temperature $T=10$ MeV and $Y_p=0.5$. 
With increasing $P$, the overlap of the wave function with the occupied phase space
becomes smaller so that the Pauli blocking mechanism becomes less efficient, and the shift is decreasing. 
A Gaussian fit as used in Refs. \cite{Typel,R} is reasonable at small densities. 
The merger with the continuum leads to deviations so that, in principle, higher orders of $P^2$
can be considered in the dispersion relation. 

To give a more general fit that reproduces the exact solution within a few percent within the entire $P$ space, we take
\begin{eqnarray}
\Delta E_d^{\rm Pauli}(P;T,n,Y_p)&=&\Delta E_d^{\rm Pauli}(0;T,n,Y_p) \frac{e^{-g_{a,d}(T,n) P^2/\hbar^2}}{1+g_{b,d}(T,n) P^2/\hbar^2}, 
\end{eqnarray}
for $ n \le  n_d^{\rm Mott}(T,Y_p)$,
and
for $ n>  n_d^{\rm Mott}(T,Y_p)$:
\begin{eqnarray}
\label{B7}
\Delta E_d^{\rm Pauli}(P;T,n,Y_p)&=& E_d 
\frac{e^{-g_{a,d}(T,n) (K- K_d^{\rm Mott}(T,n,Y_p))^2}}{1+g_{c,d}(T,n) (K- K_d^{\rm Mott}(T,n,Y_p))^2}\,, 
\end{eqnarray}
with the quantities
\begin{eqnarray}
 g_{a,d}(T,n)&\!\!\!\!=&\!\!\!\!(0.279145-0.0099364\, T+0.00017819 \,T^2)  \nonumber \\ 
 &&+ (0.41393+0.0069374 \,T) \,n -1.65\, n^2\,, \nonumber\\
g_{b,d}(T,n)&\!\!\!\!=&\!\!\!\!\left[ \frac{\hbar^2}{8mT}\left\{1- \frac{2 f_{d,2} f_{d,3}^2}{3T} \left[ \frac{1}{F(f_{d,3} \left(1+f_{d,2}/T)^{1/2}\right)}-1\right] \right\}-g_{a,d}(T,n)\right] \nonumber \\ 
&&\times \left[1+(25.797-10.456\, T+0.29668\, T^2) \,n \right], \nonumber\\
g_{c,d}(T,n)&\!\!\!\!=&\!\!\!\!\left[ \frac{\hbar^2}{8mT}\left\{1- \frac{2 f_{d,2} f_{d,3}^2}{3T} \left[ \frac{1}{F(f_{d,3} \left(1+f_{d,2}/T)^{1/2}\right)}-1\right] \right\}-g_{a,d}(T,n)\right] \nonumber \\ 
&&\times \left[1-(41.703-2.6883\, T+0.18684\, T^2) \,n \right]\nonumber\\&&
+(1.0613-0.083155\, T+0.0018922\, T^2) K_d^{\rm Mott}(T,n,Y_p)\,. \nonumber \\ {}
\end{eqnarray}
Large deviations of the fit and the exact solution arise for small c.o.m. momenta near the Mott density  $n_d^{\rm Mott}(T,Y_p)$, see Fig. \ref{fig:2}.

\section{\label{app3} Symmetrization of the in-medium Hamiltonian }

The in-medium Hamiltonian (\ref{waveA}) is not hermitian, $H^{\rm matter}(1\dots A,1'\dots A') \neq [H^{\rm matter}(1'\dots A',1\dots A)]^*$, 
see also \cite{Danielewicz}.
In the case $A=2$ it can be transformed to a hermitian Hamiltonian, 
$[1-f(1)-f(2)]w(1,2)w(1'2') \to [1-f(1)-f(2)]^{1/2}w(1,2)w(1'2') [1-f(1')-f(2')]^{1/2}$  
transforming also the wave functions  $ \psi_{d P}(12) \to [1-f(1)-f(2)]^{-1/2} \psi_{d P}(12)$. In the nondegenerate case where $f(1) \ll 1$,
 we can take $ [1-f(1)/2-f(2)/2]w(1,2)w(1'2') [1-f(1')/2-f(2')/2]$ for the symmetrised Hamiltonian. 
In the degenerate case, new effects such as pairing will occur. In contrast to the Pauli blocking term $[1-f(1)][1-f(2)]$
used in the Brueckner theory, where hole-hole contributions are neglected, the full Pauli blocking term $[1-f(1)-f(2)]$ leads to the 
Gor'kov equation in the BCS theory of superfluidity.












\end{document}